\documentclass[10pt,a4paper]{article}
\usepackage{graphicx,amsmath,amssymb,amsbsy,amsfonts,fixmath,isomath,theorem}

\newcommand{\R}{{\Bbb R}}

\newcommand{\D}[2]{ \ensuremath{ \frac{d #1 }{d #2 } }}

\newcommand{\vect}[1]{\ensuremath{ \mathbold #1 } }

\theoremstyle{break}
\theoremstyle{break}

\begin{document}

\title{Reduction of dynamical biochemical reaction networks in
computational biology}


 \author{O.~Radulescu$^1$, A.N.~Gorban$^2$, A.~Zinovyev$^3$ and  V.~Noel$^4$ \\
  \small  $^1$ DIMNP UMR CNRS 5235, University of Montpellier 2, Montpellier, France. \\
  \small  $^2$ Department of Mathematics, University of Leicester, LE1 7RH, UK. \\
  \small  $^3$ Institut Curie, U900 INSERM/Curie/Mines ParisTech, 26 rue d'Ulm,
F75248 Paris, France. \\
 \small $^4$IRMAR UMR 6625, University of Rennes 1, Rennes, France.
}

\maketitle



\centerline{\bf Abstract} Biochemical networks are used in computational biology, to
model the static and dynamical details of systems involved in cell signaling, metabolism,
and regulation of gene expression. Parametric and structural uncertainty, as well as
combinatorial explosion are strong obstacles against analyzing the dynamics of large
models of this type. Multi-scaleness is another property of these networks, that can be
used to get past some of these obstacles. Networks with many well separated time scales,
can be reduced to simpler networks, in a way that depends only on the orders of magnitude
and not on the exact values of the kinetic parameters. The main idea used for such robust
simplifications of networks is the concept of dominance among model elements, allowing
hierarchical organization of these elements according to their effects on the network
dynamics.  This concept finds a natural formulation in tropical geometry. We revisit, in
the light of these new ideas, the main approaches to model reduction of reaction
networks, such as quasi-steady state and quasi-equilibrium approximations, and provide
practical recipes for model reduction of linear and nonlinear networks. We also discuss
the application of model reduction to backward pruning machine learning techniques.

\section{Introduction}

During the last decades, biologists have identified a wealth of molecular components and
regulatory mechanisms underlying the control of cell functions. Cells integrate external
signals through sophisticated signal transduction pathways, ultimately affecting the
regulation of gene expression, including that of the signaling components. Metabolic
functions are sustained and controlled by complex machineries involving genes, enzymes
and metabolites. The genetic regulations result from the coordinate effect of many,
mutually interacting genes. These regulations involve many molecular actors, including
proteins and regulatory RNAs, which form large, intricate networks.

Current dynamical models of cellular molecular processes are small size networks. These
small scale models, that are subjective simplifications of reality, can not take into
account the specificities of regulatory mechanisms. New methods are needed, allowing to
reconcile small scale dynamical models and large scale, but static, network
architectures. The main obstacle to increasing the size of dynamical networks is the
incomplete information, on the parameters and on the mechanistic details of the
interactions. In vivo values of the parameters depend on crowding and heterogeneity of
the intracellular medium, and can be orders of magnitude different from what is measured
in vitro. Furthermore, learning models from data suffer for non-identifiability and
over-fitting problems. Thus, model reduction is an avoidable step in the study of large
networks, allowing to extract the essential features of the model, that can then be
identified from data.
Model reduction in computational biology should have several particularities.

First of all, model reduction should cope with  parametric incompleteness and/or uncertainty.

A certain class of reduction methods are parameter independent and automatically comply
with this specificity.
In biochemical networks, the number of possible chemical species
grows combinatorially due to numerous possibilities of interactions between molecules
with multiple interaction sites.
The exact lumping methods \cite{Borisov05,Conzelmann06} reduce the number of microstates
and avoid combinatorial explosion in the description and analysis of large models of
receptor and scaffold signalling. A similar technique \cite{feret2009internal} is used to
rationally organize supramolecular complexes in rule-based modeling \cite{danos2007rule}
of biochemical networks. Other, parameter independent, coarse-graining techniques are
graphical methods formalizing node deletion and merging operations in biochemical
networks \cite{gay2010graphical}, pooling of metabolites in large scale metabolic
networks \cite{papin2004hierarchical,jamshidi2008formulating}, or extensive searches in
the set of all possible lumps \cite{dokoumetzidis2009proper}. Finally, qualitative
reduction methods were used to simplify large logical regulatory graphs, adequately
suppressing nodes and defining sub-approximating dynamics
\cite{naldi2009reduction,naldi2011dynamically}.

Secondly, biochemical processes governing network dynamics span over many timescales. For
example, changing gene expression programs can take hours and even days while protein
complex formation goes on the second scale and post-translational protein modifications
take minutes to happen. Protein life half-times can vary from minutes to days. Model
reduction should exploit multiscaleness. Asymptotic dynamics of networks with slow and
fast processes, can be strongly simplified using various ideas such as inertial and
invariant manifolds (IM) and averaging approximations.

The iterative methods of IM aim to find a slow low dimensional IM, containing the
asymptotic dynamics \cite{GorbanKarlin1994,GorbanKarlin03,Roussel91}. The Computational
Singular Perturbation (CSP) \cite{lamm,GorbanChiavazzo07} aims to find even more, the
slow IM and, in addition, the geometry of fast foliation. Invariant manifolds can be
calculated by various other methods
\cite{gorban_book,Grids,Roussel91,KazantzisGood2002,Krauskopf05}.

Very popular are the methods for computation of a ``first approximations" to the slow IM.
The classical quasi steady-state approximation (QSS) was proposed by
\cite{bodenstein1913theorie} and was elaborated into an important tool for analysis of
chemical reaction mechanism and kinetics \cite{semenoff1939kinetics,
christiansen1953elucidation, helfferich1989systematic}. The classical QSS is based on the
relative smallness of concentrations of some of "active" reagents (radicals,
concentration of enzyme and substrate-enzyme complexes or amount of active centers  on
the catalyst surface) \cite{aris1965introduction, segel1989quasi,Yablonskiiatal1991}. The
quasiequilibirium approximation (QE) has two basic formulations: the thermodynamic
approach, based on conditional entropy maximum (or free energy conditional minimum), or
the kinetic formulation, based on equilibration of fast reversible reactions.  The very
first use of the entropy maximum dates back to Gibbs \cite{gibbs2010elementary}.
Corrections to QE approximation with applications to physical and chemical kinetics were
developed by \cite{gorban2001corrections,gorban_book}. An important, still unsolved,
problem of these two approximations is the detection of QSS species and QE reactions
without application of all machinery of the IM or CSP methods. Indeed, not all reactions
with large constants are at quasi-equilibrium, and there are no simple rules to find QSS
species if there is no such hints as a small amount of a conserved quantity (like the
total concentration of enzyme). The method of Intrinsic Low Dimensional Manifolds (ILDM)
\cite{ILDM1992,BykovMaas2006} provides an approximation of a low dimensional invariant
manifold and works as a first step of CSP \cite{KaperKaper2002}.

Another method allowing to simplify multiscale dynamics is averaging. This idea can be
tracked back to Poincar\'e's perturbative treatment of the many body problem in celestial
mechanics \cite{poincare_book}, further developed in classical mechanics by other authors
\cite{arnold_book,lochack_meunier}, and also known as adiabatic or Born-Oppenheimer
approximation in quantum mechanics \cite{messiah_book}. Rather generally, averaging can
be applied when some fine scale variables of the system are rapidly oscillating. Then,
the dynamics of slow, coarse scale variables, can be obtained by time averaging the
system over a timescale much larger than the period of the fast oscillations. The way to
perform averaging, depends on the structure of the system, namely on the definition of
the coarse grained and fine variables \cite{bogo61,artstein96,
acharya2006computational,sawant2006model,acharya2010coarse,givon04,slemrod2011averaging}.

Some of these ideas have been implemented in computational biology tools. Systems biology
markup language SBML \cite{hucka2003systems} can allocate a "fast" attribute to reaction
elements. Fast reaction specification can be taken into account by computational biology
softwares such as VirtualCell \cite{slepchenko2003quantitative} that implements a QE
approximation algorithm \cite{slepchenko2000numerical}. Similarly, the simulation tool
COPASI \cite{hoops2006copasi} implements the ILDM method \cite{surovtsova2009accessible}.

Finally, multiscaleness does not uniquely apply to timescales but equivalently to
abundances of various species in these networks. mRNA copy numbers can change from some
units to tens of thousands, and the dynamic concentration range of biological proteins
can reach up to five orders of magnitude. Furthermore, the DNA molecule has only one or a
few copies. Low copy numbers lead, directly or indirectly (a species can be stochastic
even if present in large copy numbers), to stochastic gene expression. In computational
biology, model reduction should thus cope not only with deterministic, but also with
stochastic and hybrid models. The need to reduce large scale stochastic models is acute.
Indeed, stochastic simulation algorithm (SSA, \cite{gillespie76,gillespie1977cvs}) can be
very expensive in computer time when applied to large unreduced models, precluding model
analysis and identification. For this reason, extensive effort has been dedicated to
adapting the main ideas used for model reduction of deterministic models, namely exact
lumping, invariant manifolds, QSS, QE, and averaging, to the case of stochastic models.

Reduction of stochastic rule-based models, based on a weakened version of the exact
lumpability criterion, has been proposed by \cite{feret2010stochastic} to define abstract
species or stochastic-fragments that can be further used in simplified calculations.
Multiscaleness of stochastic models is two-fold, it affects both species and reaction
rates. This has been exploited in hybrid stochastic simulation schemes that are, for the
most of them, based on a partition of the biochemical reactions in fast and slow
reactions \cite{haseltine,burrage2004multi,alfonsi2005ash,haseltine2005origins,
alfonsi,salis2005accurate,kaznessis2006multi,harris2006partitioned,surovtsova2006acr,salis2006mhh,griffith2006dph,
ball06,li2008algorithms,gomez2008enhanced,pahle2009biochemical}. Conversely, mixed
partitions, using both reactions and species can exploit both types of multiscaleness and
more appropriately unravel a rich variety of stochastic functioning regimes such as
piece-wise deterministic, switched diffusions, diffusions with jumps, as well as averaged
processes \cite{TSI,crudu2009hybrid,crudu2011convergence} only partially covered by some
situations discussed in \cite{mastny2007tcq}.

Machine learning approaches to parameter identification \cite{golightly2011bayesian}
could profit from Fokker-Planck  approximations, also known as diffusion approximations
or Langevin approach, of the master equation describing dynamics of stochastic networks.
Traditional approaches such as central limit theorem \cite{gillespie00,melykuti2010fast},
the $\Omega$ and the Kramers-Moyal expansions \cite{TSI,crudu2009hybrid} where used to
derive diffusion approximations. Alternatively, \cite{erban2006gene} propose diffusion
approximations for slow/fast stochastic networks, in which the drift and diffusion
parameters are obtained numerically. By the ergodic theorem, time averaging of multiscale
stochastic models boils down to a QE assumption for the fast variables. This idea has
been used in \cite{crudu2009hybrid} to reduce stochastic networks. A few computational
biology tools implement stochastic approximations \cite{salis2006mhh}.

With the exception of the parameter independent methods, all the model reduction methods
described above need a full parametrization of the model. This is a stringent
requirement, and can not be easily bypassed. Indeed, the reduction has a local validity.
The elements defining a reduced model such as IM, QSS species, QE species, depend on the
model parameters and also on the position on a trajectory of the dynamics. What one can
expect is that model reduction is robust, i.e. a given reduced model provides an accurate
approximation of the dynamics of the initial model for a wide range of parameters and
variables values. One can show that this property is satisfied by biochemical networks
with separated constants, because in this case the simplified networks depend on the
order relations among model parameters and not on the precise values of these parameters
\cite{gorbanradulescu08,radulescu2008robust,noel2011}.



The purpose of this review is not the exhaustive description of all the reduction methods that
we have delineated. We will revisit the fundamental concepts of model reduction in the light of
a new program, that should, in the long term, lead to a new generation of reduction tools
satisfying all the specific requirements of computational biology. Due to space limitations,
we restrict ourselves to deterministic models.

\section{Deterministic dynamical networks}

To construct a dynamic reaction network we need the list of
components, $\mathcal{A} =\{ A_1, ...\, A_n\}$ and the list of
reactions (the reaction mechanism):
\begin{equation}\label{stoi}
\sum_i \alpha_{ji}A_i \rightleftharpoons \sum_{k} \beta_{jk} A_k,
\end{equation}
where $j \in [1,r]$ is the reaction number.


Dynamics of nonlinear networks in homogeneous isochoric systems (fixed volume) is
described by a system of differential equations:

\begin{equation}
\D{c}{t} = P(c) =  \sum_{j=1}^r
\nu_j (R^{+}_j(c) - R^{-}_j(c))
\label{nonlinear}
\end{equation}
$c \in \R^n$ is the concentration vector, $\nu_j=\beta_j - \alpha_j$ is the global
stoichiometric vector. The reaction rates $R^{+/-}_j(c)$ are non-linear functions of the
concentrations. For instance, the mass action law reads $R^+_j(c) = k_j^+
 \prod_{i} c_i^{\alpha_{ji}}$, $R^-_j(c) = k_j^-
 \prod_{i } c_i^{\beta_{ji}}$, in which case $P_i(c)$ is a multivariate polynomial
 on the concentrations $c_j$.

\section{Multi-scale reduction of monomolecular reaction networks}

Monomolecular reaction networks are the simplest reaction networks. The structure of
these networks is completely defined by a digraph, in which vertices correspond to
chemical species $A_i$, edges correspond to reactions $A_i \to A_j$ with kinetic
constants $k_{ji} > 0$.

The kinetic equation is
\begin{equation}\label{kinur}
\D{c_i}{ t}=\sum_{j } k_{ij} c_j - \left(\sum_{j }
k_{ji}\right)c_i,
\end{equation}
or in matrix form: $\dot{c} = Kc $.

The solutions of \eqref{kinur} can be expressed in terms of left and right eigenvectors
of the kinetic matrix $K$:
\begin{equation} \label{solkinur}
c(t)= (l^0, c(0)) + \sum_{k=1}^{n-1} r^k (l^k, c(0))
\exp(-\lambda_k t)
\end{equation}
where $K  r^k = \lambda_k r^k$, and $l^k K   = \lambda_k l^k$.

Each eigenvalue $\lambda_k$ is the inverse of a timescale of the network. A reduced
network having solutions of the type \eqref{solkinur}, with eigenvectors $r^k$, $l^k$,
and eigenvalues $\lambda_k$ approximating the eigenvectors and the eigenvalues of the
original network is called a {\em multiscale approximation}.

We say that the network constants are totally separated if for all $(i,j)\neq (i',j')$
one of the relations $k_{ji} << k_{j'i'}$, or $k_{ji} >> k_{j'i'}$ is satisfied.

It was shown in \cite{gorbanradulescu08,radulescu2008robust,gorban2009asymptotology} that
the multiscale approximations of arbitrary monomolecular reaction networks with totally
separated constants are acyclic (have no cycles), and deterministic (have no nodes from
which leave more than one edge) digraphs.

In order to reduce a network with total separation, one needs only qualitative
information on the constants. More precisely, each edge of the reaction digraph can be
labeled by a positive integer representing the rank of the reaction parameter in the
ordered series of parameter values, the largest parameter (the quickest reaction) having
the lowest label. These integer labels also indicate the timescales of the processes
modeled by the network reactions.

\begin{figure}
\begin{center}
\includegraphics[width=150mm]{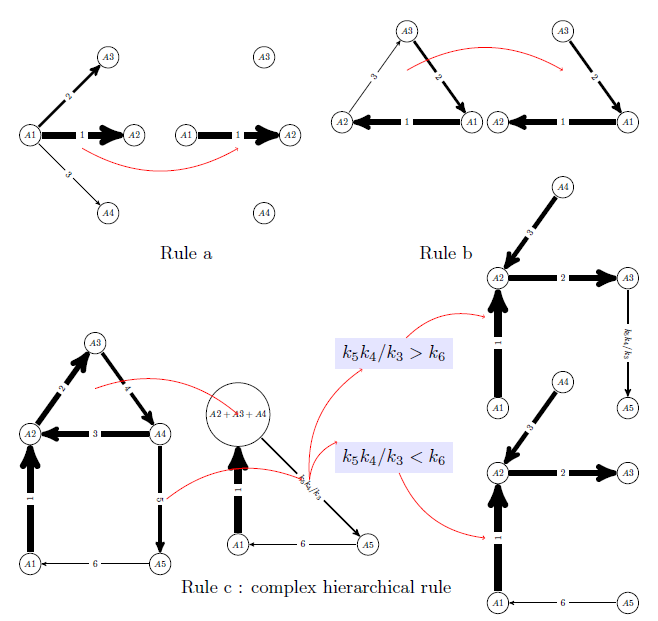}
\end{center}
\caption{ \label{fig1} A monomolecular network with total separation can be represented as a digraph with integer labels
(the quickest reaction has label 1). Two simple rules allow to eliminate competition between reactions (rule a)
and transform cycles into chains (rule b). Rule b can not be applied to cycles with outgoing slow reactions, in which case more
complex, hierarchical rules should be applied (rule c). In the rule c, first
the cycle $A_2 \to A_3 \to A_4 \to A_2$ is ``glued'' to a new node (pool $A_2+A_3+A_4$) and
the constant of the slow outgoing reaction renormalized to a monomial $k_5k_5/k_3$. Rule b is  applied to the resulting network, which is a cycle with no outgoing reactions.
The comparison of the constants $k_5k_5/k_3$ and $k_6$ dictates where this cycle is cut.
Finally, the glued cycle is restored, with its slowest step removed.}
\end{figure}

The reduced network is not always a subgraph of the initial graph. It is obtained from
this integer labeled digraph by graph re-writing operations, that can be generically
described as pruning and pooling. Two types of pruning operations are of primary
importance (see also Figure \ref{fig1}) :

\begin{description}
\item[Rule a)] If one has one node from which leave more than one edge, then all the
    edges are pruned with the exception of the fastest one (lowest integer label).
    This operation corresponds to keeping the dominant term among the terms $c_i
    k_{ij}$ consuming a species $A_i$, and reduces the node outdegree to one. The
    same principle can not be applied to reduce the indegree, because which
    production term is dominant among $k_{ij} c_j$ depends not only on $k_{ij}$ but
    also on the concentrations $c_j$.
\item[Rule b)] Cycles with separated constants can be transformed into chains, by
    elimination of the slowest step. This can be justified intuitively by topology,
    because any two nodes of a cycle are connected by two paths, one containing the
    slowest step and the other one not containing the slowest step. The latter
    shortcuts the former.
\end{description}

However, a combination of rules a) and b) is not allowed to prune slow reactions leaving cycles and further transform the cycles into chains. Indeed, the total mass of such cycles is slowly decaying because of outgoing reactions. Pruning the slow reactions that leave a cycle would keep the total cycle mass constant and produce the wrong long time approximation. In this case,
 pooling operations are needed:
\begin{description}
\item[Rule c)] Glue each cycle in the pruned system into a new vertex  and transform
    the network of {\em all initial reactions} into a new one. The concentration of
    this new component is the sum of the concentration of the glued vertices.
    Reactions to the cycles transform into reactions to the correspondent new
    vertices (with the same constants). To transform the reactions from the cycles,
    we have to calculate the normalized quasi-stationary distributions inside each
    cycle (with unit sum of the concentrations in each cycle). Let for the vertex
    $A_i$ from a cycle this concentration be $c_i^{\circ}$. Then the reaction $A_i
    \to A_j$ with the constant $k_{ji}$ transforms into the reaction from the new
    (``cycle") vertex with the constant $k_{ji} c_i^{\circ}$. The destination vertex
    of this reaction is  $A_j$ if it does not belong to a cycle of the pruned system,
    it is the correspondent glued cycle if it includes $A_j$ and does not include
    $A_i$ and the reaction vanishes if both $A_i$ and $A_j$ belong to the same cycle
    of the pruned system.
\end{description}

After pooling we have to prune (Rule a) and so on, until we get an acyclic pruned system.
Then the way back follows: we have to restore cycles and cut them (Rule b).

\begin{figure}
\begin{center}
\includegraphics[width=150mm]{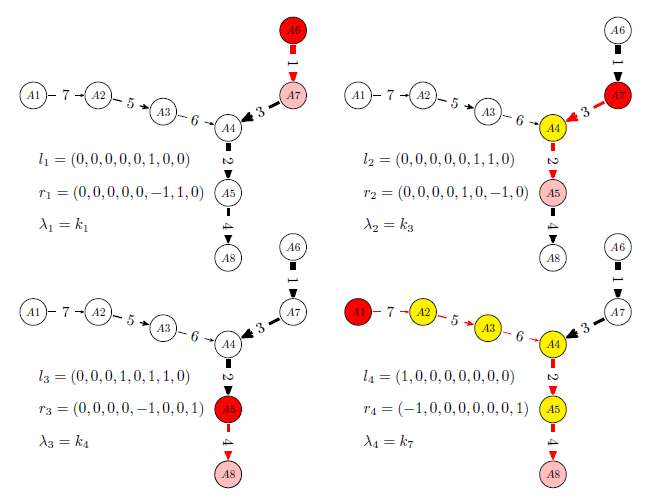}
\end{center}
\caption{\label{fig2} For a given timescale, monomolecular networks with total separation behave as a
single step: the concentrations of some species (white)
are practically constant, some
species (yellow) are rapid , low concentration, intermediates, one species (red) is gradually
consumed and another (pink) is gradually produced. We have represented the
sequence of one step approximations of a reduced, acyclic, deterministic digraph,
from the quickest time-scale $t_1=\lambda_1^{-1}$ to the
slowest one $t_4=\lambda_4^{-1}$. These one step
approximations are activated when mass is introduced at $t=0$ via the ``boundary nodes''
$A_1$ and $A_6$. }
\end{figure}

In more detail, the graph re-writing operations, are described in the Appendix and
illustrated in Figure \ref{fig1}.
The dynamics of reduced acyclic deterministic digraphs follows from their topology and
from the timescale labels. First of all, let us notice that the network has as many
timescales as remaining edges in the reduced digraph. The computation of eigenvectors of
acyclic deterministic digraphs is straightforward
\cite{gorbanradulescu08,radulescu2008robust,gorban2009asymptotology}. For networks with
total separation, these eigenvectors satisfy, in the first approximation, a $0-1$ type
property, the coordinates of $l^k$, $r^k$ belong to the sets $\{0,1\}$, and $\{0,1,-1\}$
respectively. The $0-1$ property of eigenvectors has a non-trivial consequence. On the
timescale $t_k = (\lambda_k)^{-1}$, the reduced digraph behaves as an effective reaction
(single step approximation). The effective reaction receives (from reactions acting on
smaller timescales) the mass coming from the species with coordinate $1$ in $l^k$ (pool)
and transfers it (during a time $t_k$) to the species with coordinate $1$ in $r^k$. The
successive single step approximations of an acyclic deterministic digraph are illustrated
in Figure \ref{fig2}.

Monomolecular networks with separation represent instructive examples where reduction
and qualitative dynamics result from the network topology and from the orders of
magnitude of the kinetic constants. This type of models can be used in
computational biology to reduce linear subnetworks or even binary reactions
for which one reactant is present in much larger quantities than the other (pseudo-monomolecular
approximation).

As argued by a few authors, total separation could be a generic property of
biochemical
networks \cite{furusawa2003zipf}. This property can be checked empirically by investigating the
distribution of network timescales in logarithmic scale. Whenever one finds
distributions with large support in logarithmic scale
(a log-uniform distribution is equivalent to the Zipf law, i.e. a power law distribution
with exponent $-1$, well known in critical
systems \cite{furusawa2003zipf}) total separation is valid and the above reduction method applies.

\section{Separation, dominance, and tropical geometry}

The previously presented algorithm is based on the idea of dominance, which occurs at
many levels. For instance, when several reactions compete for the same pool, all can be
pruned, excepting the dominant one (Rule a)). This simple idea is widely spread, and
corresponds to max-plus algebra: the sum of positive, well separated terms, can be
replaced by the maximum term. Max-plus algebra, that found many applications to dynamical
systems \cite{cohen1999max,van2006modelling,aubin2010macroscopic}, belong to the new
mathematical field of tropical geometry \cite{pachter2004tropical}. Tropical geometry
offers convenient solutions to solve systems of polynomial equations  with separated
monomials, to simplify and hybridize systems of polynomial or rational ordinary
differential equations with separated monomials. We can conveniently use tropical
geometry concepts to rationalize many model reduction operations and find new ones.


The logarithmic transformation $u_i = log x_i,\, 1 \leq i \leq n$, well known for drawing
graphs on logarithmic paper, plays a central role in tropical geometry
\cite{viro2008sixteenth}.

Let us consider multivariate monomials $M( \vect{x}) = a_{\alpha} \vect{x}^\alpha$, where
$\vect{x}^\alpha = x_1^{\alpha_1} x_2^{\alpha_2} \ldots x_n^{\alpha_n}$. Monomials with
positive coefficients $a_{\alpha}>0$, become linear functions, $log M = log a_{\alpha} +
<\alpha,log(\vect{x})>$, by this transformation.

There is a straightforward way to use the logarithmic transformation from tropical
geometry in order to obtain approximations of dynamical networks of the type
\eqref{nonlinear}. Let us suppose that reaction rates are polynomial functions of the
concentrations (this is satisfied by mass action law and obviously, also by monomolecular
networks), such that $\sum_{j=1}^r \nu_j (R^{+}_j(c) - R^{-}_j(c)) = \sum_{\alpha \in A}
a_{\alpha} \vect{c}^\alpha$.

We call tropicalization  of the smooth ODE system \eqref{nonlinear} the following
piecewise-smooth system:

\begin{equation}
\D{c_i}{t} = s_i exp [max_{\alpha \in A_i} \{ log( |a_{i,\alpha}|) +
< \vect{c} , \alpha > \}], \label{fraction}
\end{equation}

\noindent where $\vect{u} = (log c_1,\ldots,log c_n)$, $s_i = sign(a_{i,\alpha_{max}}) $
and $a_{i,\alpha_{max}},\, \alpha_{max}\in A_i$  denotes the coefficient of a monomial
for which the maximum occurring in \eqref{fraction} is attained.

The tropicalization associates to a polynomial $\sum_{\alpha \in A} a_{\alpha}
\vect{c}^\alpha$, the max-plus polynomial
$$P^\tau(\vect{c}) = exp [max_{\alpha \in A} \{log( |a_{\alpha}|) + < log(\vect{c}), \alpha > \}].$$

\begin{figure}
\includegraphics[width=50mm]{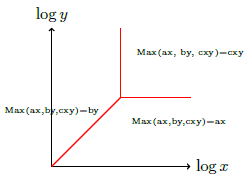}
\includegraphics[width=100mm]{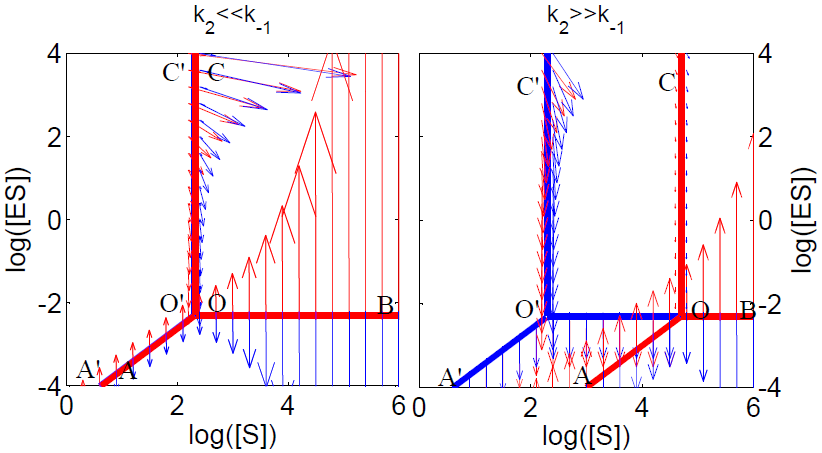}

{\hskip2truecm  a) \hskip8truecm  b)}
\caption{\label{fig3} a) The tropical manifold of the polynomial $ax + by + cxy$
on ``logarithmic paper'' is a three lines tripod.
b) The tropical manifolds for the species ES (in red) and S (in blue) for the
Michaelis-Menten mechanism. The tropicalized flow is also represented on both
sides of the tropical manifolds (with arrows, red on one side, blue on the other
side). Sliding modes correspond to blue and red arrows pointing in opposite directions.
}
\end{figure}

In other words, a polynomial is replaced by a piecewise smooth function, equal to the
largest, in absolute value, of its monomials. Thus, \eqref{fraction} is a piecewise
smooth model \cite{SASB2011} because the dominating monomials in the max-plus polynomials
can change from one domain to another of the concentration space.
The singular set where at least two of the monomials are equal, and where the max-plus
polynomial $P^\tau(\vect{c})$  is not smooth is called tropical variety
\cite{mikhalkin2007tropical}. On logarithmic paper, the tropical varieties of various
species define polyhedral domains inside which the dynamics is defined by monomial
differential equations (Figure \ref{fig3}).
Tropicalized systems remind of, but are not equivalent to, Savageau's S-systems
\cite{savageau1987recasting} that have been used for modeling metabolic networks.
S-systems are smooth systems such that the production and consumption terms of each
species are multivariate monomials. Tropicalized systems are S-systems locally, within
the polyhedral domains defined by the tropical varieties, and also along some parts of
the tropical variety (that carry sliding modes, see next section).

The tropicalization unravels an important property of multiscale systems, that is to have
different behavior on different timescales. We have seen that, on every timescale,
monomolecular networks with total separation behave like a single reaction step. This is
akin to considering only the dominant processes in the network and implies that the
tropicalization is a good approximation for monomolecular networks with total separation.
In the next section we discuss other, more general situations, that include nonlinear
networks, when the tropicalization represents an useful approximation of the smooth
dynamics.

\section{Quasi-steady state and Quasi-equilibrium, revisited}

Two simple methods for model reduction of nonlinear models with multiple timescales: the
quasi-equilibrium (QE) and the quasi-steady state (QSS) approximations. As discussed in
\cite{gorban2009asymptotology,gorban2011MM}, these  two approximations are physically and dynamically
distinct. In order to understand these differences let us refer to the simple example of
the Michaelis-Menten mechanism,

\begin{equation}
S + E \underset{k_{-1}}{ \overset{k_{1}}{\rightleftharpoons}} ES  \overset{k_2}{\rightarrow}
P + E
\end{equation}

The QSS approximation, proposed for this system by Briggs and Haldane, considers that the
total concentration of enzyme, $[E]+[ES]$ is much lower than the total concentration of
substrate, therefore complex $ES$ is a {\em low concentration, fast species.} Its
concentration is driven by concentration of $S$, hence, the simplified mechanism
correspond to pooling the two reactions of the mechanism into a unique irreversible
reaction $ S \overset{ R(S,E_{tot}) }{\longrightarrow} P $, which means that $\D{[P]}{t}
= - \D{[S]}{t} = k_2 [ES]_{QSS}$. The QSS value of the complex concentration results from
the equation  $k_1 [S] ([E]_{tot} - [ES]_{QSS}) = (k_{-1} + k_2) [ES]_{QSS}$. From this
follows that $R([S],[E]_{tot}) = k_2 [E]_{tot}[S] / (k_m + [S])$, where $[E]_{tot}$ is
the total enzyme concentration, and $k_m=(k_{-1} + k_2)/k_1$.

The QE approximation considers that the first reaction of the mechanism is a {\em  fast, reversible reaction.}
The simplified mechanism corresponds to a pooling of species.
Two pools $[S]_{tot} = [S] + [ES]$, and $[E]_{tot} = [E] + [ES]$ are conserved by the fast reversible reaction,
but only one, $[E]_{tot}$ is conserved by the two reactions of the mechanism. The pool $[S]_{tot}$
is slowly consumed by the second reaction and represents the slow variable of the system.
The single step approximation reads
$ S_{tot}  \overset{R([S]_{tot},[E]_{tot})}{\longrightarrow} P $, or equivalently
$\D{[P]}{t} = - \D{[S]_{tot}}{t} = k_2 [ES]_{QE}$. The QE value of the complex concentration
is the unique positive solution of the quadratic equation
$k_1 ([S]_{tot} - [ES]_{QE}) ([E]_{tot} - [ES]_{QE}) = k_{-1}  [ES]_{QE}$. From this it follows
that $R([S]_{tot},[E]_{tot}) = 2 k_2 [E]_{tot}[S]_{tot} ([E]_{tot} + [S]_{tot} + k_{-1}/k_1)^{-1}
(1+\sqrt{1 -4[E]_{tot}[S]_{tot}/([E]_{tot} + [S]_{tot} + k_{-1}/k_1)^2})^{-1}$. When the concentration
of enzyme is small, $[E]_{tot} << [S]_{tot}$, we obtain the original equation of Michaelis and Menten,
$R([S]_{tot},[E]_{tot}) \approx k_2 \frac{ [E]_{tot}[S]_{tot}}{k_{-1}/k_1 + [S]_{tot}}$.

One of the main difficulties to applying QE or QSS reduction to computational biology
models is that QE reactions and QSS species should be specified a priori. For some
models, biological information can be used to rank reactions according to their rates.
For instance, one knows that metabolic processes and post-transcriptional modifications
are more rapid than gene expression. However, this information is rather vague. In
detailed gene expression models some processes can be rapid, while others are much
slower. Furthermore, the relative order of these processes can be inverted from one
functioning regime to another, for instance the binding and unbinding rates of a repressor to DNA, can
be slow or fast depending on various conditions.
Even if some numerical approaches such as iterative IM, CSP and ILDM propose criteria for
detecting fast and slow processes, at present there is no general direct method to
identify QE reactions and QSS species.

Here we present two methods, based, the first one on singular perturbations, and the
second on tropical geometry ideas,  allowing to detect QE reactions and QSS species.

The first method uses simulation of the trajectories, therefore it can only be applied to
a fully parametrized model. However, in systems with separation, the sets of QE reactions
and QSS species are robust, ie remain the same for broad ranges of the parameters. One
can use imprecise parameters (resulting for instance from crude estimates or fitting) to
compute these sets. The method starts by detecting {\em slaved species}. Given the
trajectories $\vect{c}(t)$ of all species, the imposed trajectory of the $i$-th species
is a real, positive solution $c_i^*(t)$ of the polynomial equation
\begin{equation} \label{qsseq}
P_i(c_1(t),\ldots,c_{i-1}(t),c_i^*(t),c_{i+1}(t),\ldots,c_n(t))=0,
\end{equation}
where
$P_i$ is the $i$-th component of the rhs of \eqref{nonlinear}.
We say that a species $i$ is slaved if the distance between
the trajectory $c_i(t)$ and some imposed trajectory $c_i^*(t)$ is small for some time interval $I$,
$sup_{t \in I} |log(c_i(t))-log(c_i^*(t))| < \delta$, for some $\delta>0$ sufficiently small.
The remaining species, that are not slaved, are called slow species.

\begin{figure}
\begin{center}
\includegraphics[width=150mm]{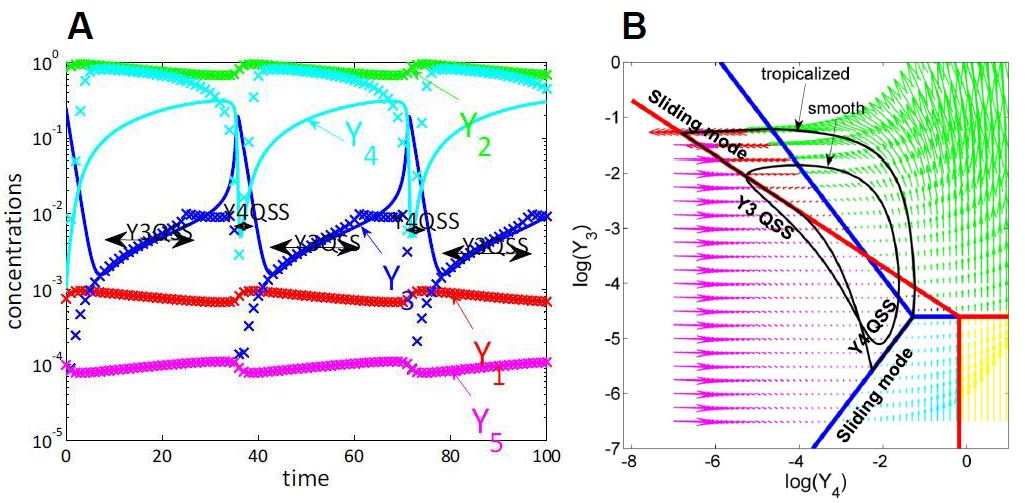}
\end{center}
\caption{\label{fig4}
Model reduction and tropicalization of a 5 variables cell cycle model defined
by the differential equations
$y_1' =k_9 y_2 - k_8 y_1 + k_6 y_3$, $y_2' =k_8 y_1 - k_9 y_2 - k_3 y_2 y_5$,
$y_3' =k_4' y_4 + k_4 y_4 y_3^2/C^2 - k_6 y_3$, $y_4' = - k_4' y_4 - k_4 y_4 y_3^2/C^2 + k_3 y_2 y_5$,
$y_5' = k_1 - k_3 y_2 y_5$, proposed in \cite{tyson1991modeling}.
(A) Comparison of trajectories and imposed trajectories show that variables
$y_1$, $y_2$, $y_5$ are always slaved, meaning that the trajectories are
close to the 2 dimensional hyperplane defined by the QE condition $k_8 y_1 = k_9 y_2$,
the QSS condition $k_1 = k_3 y_2 y_5$ and the conservation law
$y_1+y_2+y_3+y_4 = C$. The variables $y_3$, $y_4$
are slaved and the corresponding species are quasi-stationary on intervals. This means that the
dimensionality of the dynamics is further reduced to 1, on intervals.
(B) Tropicalization on logarithmic paper, in the plane of the variables $y_3$, $y_4$.
The tropical manifold consists of two tripods, represented in blue and red,
which divide the logarithmic paper into 6 polygonal sectors. Monomial vector fields defining
the tropicalized dynamics change from one polygonal domain to another. The tropicalized
(approximated) and the smooth (not reduced) limit cycle dynamics stay within bounded
distance one from another. This distance is relatively small on intervals
where the variables  $y_3$ or $y_4$ are quasi-stationary, which correspond
to sliding modes of the tropicalization.}
\end{figure}

Slaved species are rapid and are constrained by the slow species. The minimum number of variables
that we expect for a reduced model is equal to the number of slow species. The slow species
can be obtained by direct comparison of the imposed and actual trajectories. This method is
illustrated for a model of NF$\kappa$B canonical pathway in Figure \ref{fig4}.

There are two types of slaved species. Low concentration, slaved species satisfy QSS
conditions. Large concentration, slaved species are consumed and produced by fast QE
reactions and satisfy QE conditions. Because the reduction schemes are different in the
two situations, it is useful to have a method to separate the two cases. Using the values
of concentrations can work when concentrations are well separated, but may fail for a
continuum of values. A better method is to identify which are the dominant terms in the
Eq.\eqref{qsseq}. Using again the example of Michaelis-Menten mechanism, the complex ES
will be detected as slaved in both QSS and QE conditions. Eq.\eqref{qsseq} reads $k_1 [S]
[E] = (k_{-1} + k_2) [ES]$. For QE condition, the term $k_2$ will be dominated by
$k_{-1}$. We call pruned version of Eq.\eqref{qsseq} the equation obtained after removing
all the dominated monomials, in this case the equation $k_1 [S] [E] - k_{-1}  [ES] =0$.
When the pruned version is a combination of reversible reaction rates set to zero, then
the slaved species satisfy QE conditions. Again, the comparison of monomials is possible
for a fully parametrized model, however we expect this comparison to be robust for models
with separation.

The second method to identify QE and QSS conditions from the calculation of the
tropicalization \eqref{fraction}. This can be done formally and do not require simulation
of trajectories and numerical knowledge of the parameters. Indeed, is was shown in
\cite{SASB2011} that there is a relation between sliding modes of the tropicalized system
\eqref{fraction}  and the QSS or QE conditions. Sliding modes are well known for ordinary
differential equations with discontinuous vector fields \cite{filippov1988differential}.
In such systems, the dynamics can follow discontinuity hypersurfaces where the vector
field is not defined. When the discontinuity hypersurfaces are smooth and $n-1$
dimensional ($n$ is the dimension of the vector field) then the conditions for sliding
modes read:
\begin{equation}
<n_+(x), f_+(x)> < 0, \quad <n_-(x), f_-(x)> < 0, \quad x \in \Sigma,
\label{slidingmode}
\end{equation}
where $f_+,f_-$ are the vector fields on the two sides of $\Sigma$ and
$n_+= -n_-$ are the interior normals.

In \cite{SASB2011} we have shown the following. If the smooth dynamics obeys QE or QSS
conditions and if the pruned polynomial $\tilde P$ defining the fast dynamics is a
2-nomial, $\tilde P_i(\vect{c}) = a_{1} \vect{c}^{\alpha_1} + a_{2} \vect{c}^{\alpha_2}$,
then the QE or QSS equations define a hyperplane of the tropical variety of $\tilde P$,
namely $S = \{ <log(\vect{c}),\alpha_1-\alpha_2> = log (|a_{1}|/|a_{2}|) \}$. The
stability of the QE of QSS manifold implies the existence of a sliding mode of the
tropicalization \eqref{fraction} along this hyperplane. This result suggests that
checking the sliding mode condition \eqref{slidingmode} on the tropical manifold,
provides a method of detecting QE reactions and QSS species.

To illustrate this method, let us use again the Michaelis-Menten example. In this case,
two conservation laws allow elimination of two variables $E$ and $P$ and the dynamics can
be described by two ODEs:
\begin{eqnarray}
\D{[S]}{t} &=& - k_1 E_{tot}[S] + k_1 [S] [ES] + k_{-1} [ES] \notag \\
\D{[ES]}{t}& =&  k_1 E_{tot}[S] - k_1 [S] [ES] - (k_{-1} + k_2)[ES]
\end{eqnarray}
The tropical manifolds of the two species $S$ and $ES$ are tripods with parallel arms
like in Figure~\ref{fig3}. Indeed, the slopes of the arms of tropical manifold are only
given by the powers of different variables of the monomials, and these are the same for
the two species. Investigation of the flow field close to the tripod arms identifies
sliding modes on an unbounded subset $AOB$ of the tropical manifold of the species $ES$.
This subset is a global attractor of the tropicalized dynamics and represents a
tropicalized version of the invariant manifold of the smooth system. If the initial data
is not in this set, the tropicalized trajectory converges quickly to it and continues on
it as  a sliding mode. When $k_2 >> k_{-1}$, $ES$ satisfies QSS conditions leading to the
Michaelis-Menten equation. The arm $AO$ of the tropical manifold of the species $ES$
carry a sliding mode, has the equation $k_1 E_{tot}[S] = (k_{-1} + k_2)[ES] >> k_1 [S]
[ES]$, and corresponds to the linear regime of the Michaelis-Menten equation. Similarly,
the arm $OB$ of the tropical manifold of $ES$ has the equation $   k_1 E_{tot}[S] =  k_1
[S] [ES] >> (k_{-1} + k_2)[ES]$ and corresponds to the saturated regime of the
Michaelis-Menten equation. When $k_2 << k_{-1}$, the tropical manifolds of the two
species $S$ and $ES$ practically coincide. Both species are rapid and satisfy QE
conditions, namely $k_1 E_{tot}[S] = k_{-1}[ES] >> k_1 [S] [ES]$ on the arm $AO$ and $
k_1 E_{tot}[S] = k_1 [S] [ES] >> k_{-1}[ES]$ on the arm $OB$.

The tropicalization can thus be used to obtain global reductions of models. Even when
global reductions are not possible (sliding modes leave the tropical manifold or simply
do not exist), the tropicalization can be used to hybridize smooth models, ie transform
them into piecewise simpler models (modes) that change from one time interval to another.
These changes occur when the piecewise smooth trajectory of the system meets a hyperplane
of the tropical manifold and continues as a sliding mode along this hyperplane or leaves
immediately the hyperplane. Hybridization is a particularly interesting approach to
modeling cell cycle. Indeed, progression of the cell cycle is a succession of several
different regimes (phases). This strategy is illustrated in Figure~\ref{fig4} for a
simple cell cycle model.

\section{Graph rewriting for large nonlinear, deterministic, dynamical networks}

We have seen that model reduction of monomolecular networks with total separation is
based on graph rewriting operations.



Similarly, QSS and QE approximations can be used to produce simpler networks from large
nonlinear networks. The classical implementation of these approximations leads to
differential-algebraic equations. It is however possible to reformulate the simplified
model as a new, simpler, reaction network. We showed in the previous section how to do
this for the Michaelis Menten mechanism under different conditions. In general one has to
solve the algebraic equations corresponding to QE or QSS conditions, eliminate (prune)
QSS species and QE reactions, pool reactions (for QSS approximation) or species (for QE
approximation), and finally calculate the kinetic laws of the new reactions.

By reaction pooling we understand here replacing a set of reactions by a single reaction
whose stoichiometry vector $\nu$ is the sum of the stoichiometry vectors $\nu_i$ of the
reactions in the pool, $\nu = \sum_i \gamma_i \nu_i$. If the reactions are reversible
then the coefficients $\gamma_i$ can be arbitrary integers, otherwise they must be
positive integers. Reaction pools conserve certain species that where previously consumed
or produced by individual reactions in the pools. These species were called intermediates
in \cite{radulescu2008robust}. The species that are either produced or consumed by the
pools were called terminal in \cite{radulescu2008robust}. For example, an irreversible
chain of reactions $A_1 \to A_2 \to A_3$ can be pooled onto a single reaction $A_1 \to
A_3$, which in terms of stoichiometry vectors reads $\begin{bmatrix} -1 \\ 0 \\ 1
\end{bmatrix} =  \begin{bmatrix} -1 \\ 1 \\ 0 \end{bmatrix} + \begin{bmatrix} 0 \\ -1 \\
1 \end{bmatrix}$. In this example $A_1$, $A_3$ are terminal species and $A_2$ is an
intermediate species. Reaction pooling is used with QSS conditions, in which case the
intermediates are the QSS species.

By species pooling we understand replacing a set of species concentrations $\{ c_i \}$ by a linear combination
with positive coefficients of species concentrations, $\sum_i b_i c_i$. Species pooling is
used with QE conditions.

In general, the reaction and species pools result from linear algebra. Indeed, let us consider the
matrix $S^f$ that defines the stoichiometry of the rapid subsystem. For the QSS
approximation, the matrix $S^f$ has a number of lines equal to the number of QSS species.
The columns of this matrix are the stoichiometries of the reactions in the model,
restricted to the QSS species. We exclude zero valued columns, i.e. reactions that do not
act on QSS species. For the QE approximation, the number of columns of the matrix $S^f$
is equal to the number of QE reactions, and the lines of $S^f$ are the stoichiometries of
QE reactions. We exclude zero valued lines corresponding to species that are not affected
by QE reactions.

In QE conditions, species pools are defined by vectors in the left kernel of $S^f$,

\begin{equation}
b^T S^f =0
\end{equation}

The vectors $b$, that are conservation laws of the fast subsystem, define linear
combinations of species concentrations that are the new slow variables of the system. Of
course, one could eliminate from these combinations, the conservation laws of the full
reaction network, that will be constant (see Appendix).

In QSS conditions, reaction pools (also called routes) are defined by vectors in the right
kernel of  $S^f$,

\begin{equation}
S^f \gamma =0
\label{rker}
\end{equation}

According to the definition \eqref{rker}, a reaction pool does not consume or produce QSS
species (these are intermediates). One can impose, like in \cite{radulescu2008robust}, a
minimality condition for choosing the reaction pools. A reaction pool is minimal if there
is no other reaction pool with less nonzero stoichiometry coefficients. This is
equivalent to choosing reaction pools as elementary modes of the fast subsystem.

After pooling, QE and QSS algebraic conditions must be solved and the rates of the new
reactions calculated. The new rates should be chosen such that the remaining species and
pools of species satisfy the simplified ODEs. The choice of the rates is not always unique
 (some uniqueness conditions are discussed in \cite{radulescu2008robust}, see
 also the Appendix). In order
to compute the new rates, one has to solve QE and QSS equations. For network with
polynomial or rational rates, this implies solving large systems of polynomial equations.
The complexity of this task is double exponential on the size of the system
\cite{noel2011}, therefore one needs approximate solutions. Approximate solutions of
polynomial equations can be formally derived when the monomials of these equations are
well separated. Some simple recipes were given in \cite{radulescu2008robust} and could be
improved by the methods of tropical geometry.

These ideas were used in \cite{radulescu2008robust} to reduce several models of NF-$\kappa$B
signalling (Figure \ref{fig7}).

The NF-$\kappa$B activation pathway is complex at many levels. NF-$\kappa$B is
sequestered in the cytoplasm by inactivating proteins named I$\kappa$B. There are five
known members of the NF-$\kappa$B family in mammals, Rel (c-rel), RelA (p65), RelB,
NF-$\kappa$B1 (p50 and its precursor p105) and NF-$\kappa$B2 (p52 and its precursor
p100). This generates a large combinatorial complexity of dimers, affinities and
transcriptional capabilities. I$\kappa$B family comprises seven members in mammals
(I$\kappa$B$\alpha$, I$\kappa$B$\beta$, I$\kappa$B$\epsilon$, I$\kappa$B$\gamma$, Bcl-3).
All these inhibitors display different affinities for NF-$\kappa$B dimers, multiplying
the combinatorial complexity. The activation of NF-$\kappa$B upon signalling, occurs by
phosphorylation by a kinase complex, then ubiquitination, and finally degradation of
I$\kappa$B molecules. The activation signal is transmitted by several possible pathways
most of them activating the kinase IKK that modifies I$\kappa$B. In the canonical
pathway, one important determinant of IKK dynamics is the protein A20 that inhibits IKK
activation. A20 expression is controlled by NF-$\kappa$B. In order to cope with this
complexity a model containing 39 species, 65 reactions and 90 parameters was proposed in
\cite{radulescu2008robust}. Of course, not all reactions and parameters of this complex
model are important. In order to determine, in a rational and systematic way, which of
the model features are critical, we have used model reduction.

Graph rewriting was performed in a modular way, by applying the pruning and pooling rules
to tightly connected submodels of the NF-$\kappa$B network. The computation of the
reaction pools was performed using Matlab and METATOOL \cite{von2006metatool}. Using submodel
decomposition reduces the complexity of computing elementary modes and of solving large
systems of algebraic equations needed for recalculating the reaction rates.

\begin{figure}
\begin{center}
\includegraphics[width=160mm]{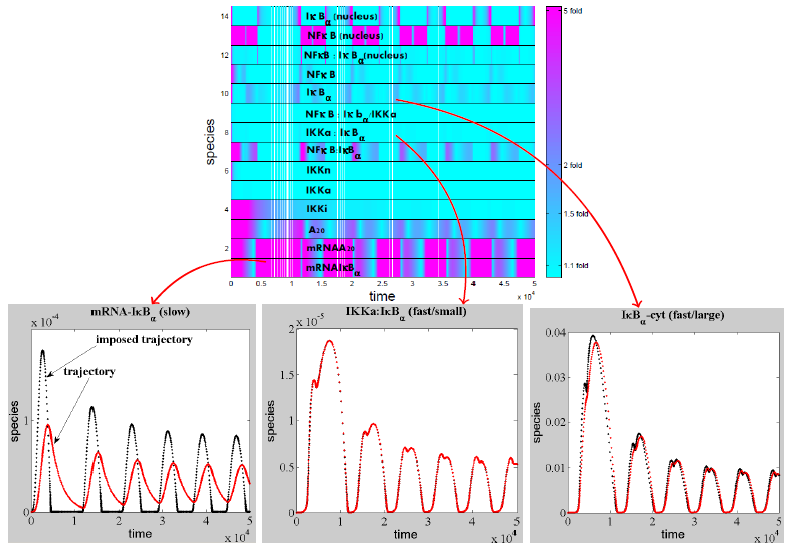}
\end{center}
\caption{\label{fig5} The ratio of the imposed and actual trajectories has been
calculated as a function of time for each species of the model
of $NF\kappa B$ canonical pathway (proposed in
\cite{lipniacki04}, model ${\mathcal M}(14,25,28)$ from \cite{radulescu2008robust}). If this ratio is close to one fold, the species is
slaved, otherwise the species is slow. Among the slaved species, some have low
concentrations and satisfy quasi-steady-state conditions, whereas other have large
concentrations and satisfy quasi-equilibrium conditions.}
\end{figure}

\begin{figure}[h!]
\begin{centering}
\includegraphics[width=160mm]{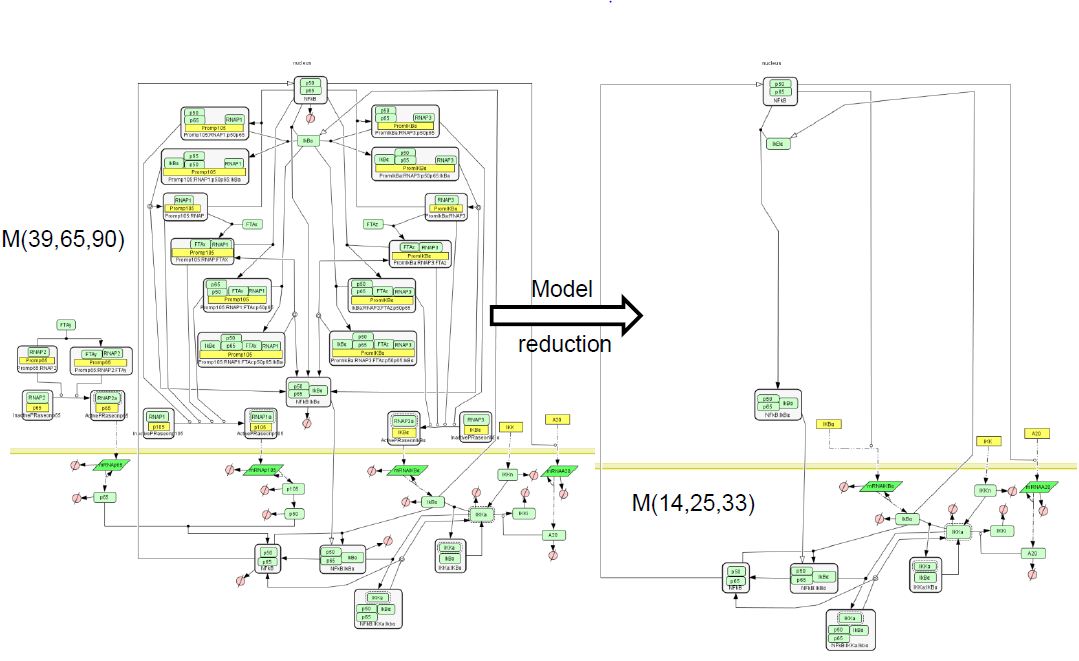}
\end{centering}
\caption{ \label{fig7}
Model of NF-$\kappa$B signaling, proposing separate production of the subunits p50, p65, the full combinatorics of their interactions as well as with the inhibitor I$\kappa$B, the positive self-regulation of p50, and in addition an A20 molecule whose production is enhanced upon NF-$\kappa$B
stimulation, and which negatively regulates the activity of the stimulus responding
kinase IKK \cite{radulescu2008robust}.
This model, denoted ${\mathcal M}(39,65,90)$  contains 39 species, 65 reactions and 90 parameters.
We have reduced it to  various levels of complexity. Among the reduced model we obtained one,
${\mathcal M}(14,25,33)$ that has the same stoichiometry
as a model published elsewhere by another author \cite{lipniacki04} and denoted
${\mathcal M}(14,25,28)$. Incidently, this is also the simplest
model in the hierarchy related to ${\mathcal M}(39,65,90)$.
The rate functions in the reduced model
are different, explaining the difference in number of parameters.
Comparison of the rate functions and of the trajectories of the
models ${\mathcal M}(14,25,33)$ and ${\mathcal M}(14,25,28)$
provided insight into the consequences of various mechanistic modeling
choices.  }
\end{figure}

To give an example of modular reduction, let us consider the set of reactions involving
six cytoplasmic located intermediates (IKK$|$active, IKK$|$inactive, IKK,
IKK$|$active:IkBa, IKK$|$active:IkBa:p50:p65, p50:p65@csl) and four terminal species
(A20, IkBa@csl, IkBa:p50:p65@csl,  p50:p65@ncl). As can be seen from Figure \ref{fig5},
the six intermediate species are slaved. The reactions of this submodel form the
cytoplasmic part of the signalling mechanism, including 11 kinase transformation
reactions, a complex release reaction, a complex formation reaction, and the NF-$\kappa$B
translocation reaction. The elementary modes of the submodel (computed using METATOOL
\cite{von2006metatool}) are used to define the reactions pools. For this submodel, we find two
elementary modes, that can be described as the modulated inhibitor degradation (IkBa@csl
$\rightarrow$ $\varnothing$), and a reaction summarizing the NF-$\kappa$B release and
translocation (IkBa:p50:p65@csl $\rightarrow$ p50:p65@ncl), respectively. In order to
compute the reaction rates of the two elementary modes as functions of the concentrations
of the terminal species, we find approximate solutions of the QSS equations for the
intermediate species and equate, for the variation rates of each terminal species, the
contributions of elementary modes to the total known variation rate in the unreduced
model (see Appendix).  The two rates are  $k_{21p1} [IkBa@csl]
[IkBa:p50:p65@csl]/((k_{21p2} + [IkBa@csl])(k_{21p3}+ [A20]))$ for the modulated
inhibitor degradation, and $k_{15p1} [IkBa:p50:p65@csl]/((k_{15p2} +
[IkBa@csl])(k_{15p3}+ [A20]))$ for the release and translocation reaction.

\section{Model reduction and model identification}

Computational biology models contain mechanistic details that can not all be
identified from available experimental data. Determining the parameters of
such complex models could lead to overfitting, describing noise, rather than
features of data, or can be simply meaningless, when model behavior
is not sensitive to the parameters. Furthermore, many model identification methods \cite{golightly2011bayesian} suffer from the
"curse of dimensionality" as it becomes increasingly difficult to cover the parameter
space when the number of parameters increases.
A rather efficient strategy to bypass these
problems is to use model reduction. This method is known in machine learning as
backward pruning or post-pruning
\cite{witten2005data}. It consists in finding a complex model that fits data well and then
prune it back to a simpler one that also fits the data well. Far from being redundant,
backward pruning can be successfully used in computational biology. Rather often, one
starts with a complex model coping with mechanistic details of the network regulation.
Then, over-fitting and problems of  identifiability of the parameters are avoided by
model reduction. By model reduction the mechanistic model is mapped onto a simpler,
phenomenological model. For instance, gene transcription and translation can be
represented as one step and one constant in a phenomenological model, but can consist of
several steps such as initiation, transcription of mRNA leading region, ribosome binding,
translation, folding, maturation, etc. in a complex model. Not all of these steps are
important for the network functioning and not all parameters are identifiable from the
observed quantities. Following reduction, the inessential steps are pruned and several
critical parameters are compacted into a few effective parameters that are
identifiable.

As discussed in \cite{eccs06,radulescu2008robust,notredame09,ferguson2012reconciling},
model reduction unravels the important features and the critical parameters
of the model.

Using model reduction for determining critical features of the model has many advantages
relative to numerical sensitivity studies
\cite{Rabitz83,gunawan2005sensitivity,ihekwaba04}:  this approach is less time consuming,
brings more insight, and is based on qualitative comparison of the order of the
parameters and therefore does not need exhaustive scans of parameter values.  In the
applications described in
\cite{eccs06,radulescu2008robust,notredame09,ferguson2012reconciling},  the critical
parameters of the pruned model are combinations (most often monomials) of the parameters
of the complex models. As only the critical combinations can be fitted from data it is
important to have estimates of some individual parameters, allowing to determine the
remaining ones.

\begin{figure}[h!]
\begin{center}
\includegraphics[width=120mm]{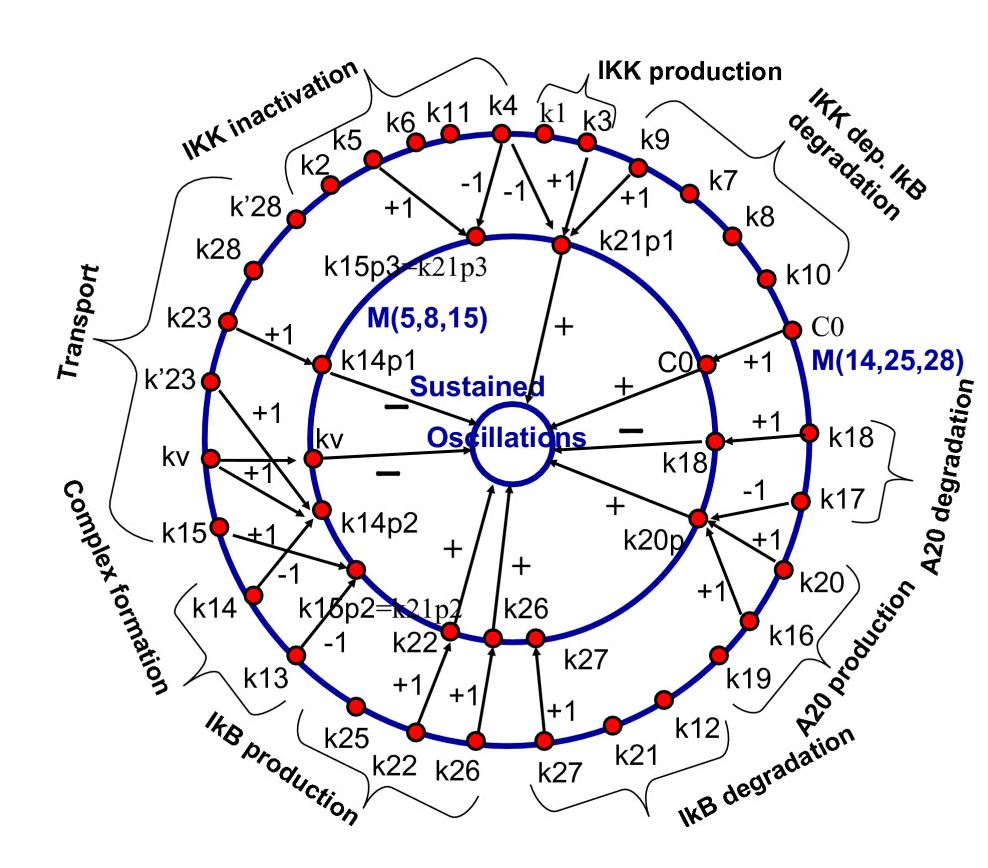}
\caption{ \label{fig8} The model ${\mathcal M}(14,25,28)$ from
from \cite{radulescu2008robust} (first proposed in
\cite{lipniacki04})
was used to
generate a hierarchy of simpler models, the simplest one being ${\mathcal M}(5,8,15)$.
We show the mapping between the parameters of the models M(14, 25, 28) and M(5, 8, 15).
Parameters of the first model are gathered into monomials that are parameters
of the reduced model. The integers on the arrows connecting parameters represent the corresponding powers of the parameters in the monomial.
The innermost circle represents a dynamical property of the model that is influenced
positively, negatively, or negligibly  by the effective parameters (parameters of the reduced model).
From \cite{radulescu2008rsm}.}
\end{center}
\end{figure}

This methodology has been first proposed in \cite{radulescu2008robust}. The model
reduction of the NF-$\kappa$B model in \cite{radulescu2008robust} leads to new, effective
parameters that are monomials of the parameters of the complex
model. The correspondence between the initial parameters and the effective parameters
is shown in Figure \ref{fig8}. Although not fully exploited in the theoretical study
\cite{radulescu2008robust}, this mapping can be used for model identification from
experimental data. Parameters of the reduced model have increased observability
and could be obtained from experimental data. The values of
the effective parameters can be used to constrain the parameters of the full model. Some of
the parameters of the full model, that are not critical or contribute to effective
parameters together with other parameters
remain arbitrary and could be fixed to generic values.

\section{Conclusion}

The mathematical techniques described in this paper define strategies for the study of
large dynamical network models in computational biology. Large networks are needed in
order to understand context dependence, specialization, and individuality of the cell
behavior. Extensive pathway database accumulation supports somehow the idea that
biological cell is a puzzle of networks and pathways, and that once these are put
together in a tightly bound, coherent map, the cell physiology should be unraveled by a
computer simulation. Actually, confronting biochemical networks with real life is not an
easy challenge. Model reduction techniques are needed to bring us one step closer to this
objective, as these methods can reveal critical features of complex organizations.

We have proposed that the ideas of limitation and dominance are fundamental for
understanding computational biology dynamical models. The essential, critical features of
systems with many separated time scales, can be resumed by a dominant, reduced,
subsystem. This dominant subsystem depends on the order relations between model
parameters or combinations of model parameters. We have shown how to calculate such a
dominant subsystem for linear and nonlinear networks. Geometrical interpretation of these
concepts in terms of tropicalization provides a powerful framework, allowing to identify
invariant manifolds, quasi-steady state species and quasi-equilibrium reactions.
We have also discussed how model reduction can be applied to backward pruning parameter learning strategies.

Future efforts are needed to extend these mathematical ideas and model reduction
algorithms and implement them into computational biology tools.


\section*{Appendix : algorithms}

\subsection*{\bf Algorithm 1 : reduction of monomolecular networks with separation}

This algorithm consists of three procedures.

{\bf I. Constructing of an auxiliary reaction network: pruning.}

For each $A_i$ branching node (substrate of several reactions) let us define $\kappa_i$
as the maximal kinetic constant for reactions $A_i \to A_j$:
$\kappa_i =\max_j \{k_{ji}\}$. For correspondent $j$ we use the notation
$\phi(i)$: $\phi(i)={\rm arg \,max}_j \{k_{ji}\}$.

An auxiliary reaction network $\mathcal{V}$ is the set of reactions obtained by keeping
only $A_i \to
A_{\phi(i)}$ with kinetic constants $\kappa_i$ and discarding the other, slower reactions.
Auxiliary networks have no branching, but they can have cycles and confluences.
The correspondent
kinetic equation is
\begin{equation}\label{auxkinet}
\dot{c}_i =  -\kappa_i c_i + \sum_{\phi(j)=i} \kappa_j c_j,
\end{equation}

If the auxiliary network contains no cycles, the algorithm stops here.

{\bf II gluing cycles and restoring cycle exit reactions}

In general, the auxiliary network $\mathcal{V}$ has several cycles $C_1,C_2, ...$ with
lengths $\tau_1, \tau_2,...>1$.

These cycles will be ``glued" into points and all nodes in the cycle $C_i$, will be
replaced by a single vertex $A^i$. Also, some of the reactions that were pruned in the
first part of the algorithm are restored with renormalized rate constants. Indeed,
reaction exiting a cycle are needed to render the correct dynamics: without them, the
total mass accumulates in the cycle, with them the mass can also slowly leave the cycle.
Reactions $A \to B$ exiting from cycles ($A \in C_i$, $B \notin C_i$) are changed into
$A^i \to B$ with the rate constant renormalization: let the cycle $C^i$ be the following
sequence of reactions $A_{1} \to A_{2} \to ... A_{\tau_i} \to A_1$, and the reaction rate
constant for $A_j \to A_{j+1}$ is $k_j$ ($k_{\tau_i}$ for $A_{\tau_i} \to A_1$). The
quasi-stationary normalized distribution in the cycle is:
$$c_j^\circ=\frac{1}{k_j}\left(\sum_{j=1}^{\tau_i} \frac{1}{k_j}\right)^{-1}\, , \; j=1, \ldots
, \tau_i\, .$$ The reaction $A_j \to B$  ($A \in C_i$, $B \notin C_i$) with the rate
constant $k$ is changed into $A^i \to B$ with the rate constant $c_j^\circ k$.

Let the cycle $C_i$ have the limiting steps that is much slower than other reactions. For
the limiting reaction of the cycle $C_i$ we use notation $k_{\lim \, i}$. In this case,
$c_j^\circ= k_{\lim \, i}/k_j$. If $A=A_j$ and $k$ is the rate constant for $A \to B$,
then the new reaction $A^i \to B$ has the rate constant $k k_{\lim \, i}/ k_j$. This rate
is obtained using quasi-stationary distribution for the cycle.

The new auxiliary network $\mathcal{V}^1$ is computed for the network with glued cycles.
Then we prune it, extract cycles, glue them, iterate until a acyclic network is obtained
$\mathcal{V}^m$, where $m$ is the number of iterations.

{\bf III Restoring cycles}

The previous procedure gives us the sequence of networks $\mathcal{V}^1, \ldots ,
\mathcal{V}^m$ with the set of vertices $\mathcal{A}^1, \ldots , \mathcal{A}^m$ and
reaction rate constants defined for each $\mathcal{V}^i$ in the processes of pruning and
gluing.

The dynamics of species inside glued cycles is lost after their gluing. A full
multi-scale approximation (including relaxation inside cycles) can be obtained by
restoration of cycles. This is done starting from the acyclic auxiliary network
$\mathcal{V}^m$ back to $\mathcal{V}^1$ through the hierarchy of cycles. Each cycle is
restored according to the following procedure:

\begin{itemize}
\item We start the reverse process from the glued network $\mathcal{V}^m$ on
    $\mathcal{A}^m$. On a step back, from the set $\mathcal{A}^m$ to
    $\mathcal{A}^{m-1}$ and so on, some of glued cycles should be restored and cut.
    On the $q$th step we build an acyclic reaction network on the set of vertices
    $\mathcal{A}^{m-q}$, the final network is defined on the initial vertex set and
    approximates relaxation of the initial networks.
\item To make one step back from $\mathcal{V}^m$ let us select the vertices of
    $\mathcal{A}^m$ that are glued cycles from $\mathcal{V}^{m-1}$. Let these
    vertices be $A^m_{1}, A^m_{2}, ...$. Each $A^m_i$ corresponds to a glued cycle
    from $\mathcal{V}^{m-1}$, $A^{m-1}_{i1} \to A^{m-1}_{i2} \to ... A^{m-1}_{i
    \tau_i} \to A^{m-1}_{i1}$, of the length $\tau_i$. We assume that the limiting
    steps in these cycles are $A^{m-1}_{i \tau_i} \to A^{m-1}_{i1}$. Let us
    substitute each vertex $A^m_i$ in $\mathcal{V}^m$ by $\tau_i$ vertices
    $A^{m-1}_{i1} , A^{m-1}_{i2}, ... A^{m-1}_{i\tau_i}$ and add to $\mathcal{V}^m$
    reactions $A^{m-1}_{i1} \to A^{m-1}_{i2} \to ... A^{m-1}_{i \tau_i}$ (that are
    the cycle reactions without the limiting step) with corresponding constants from
    $\mathcal{V}^{m-1}$.
\item
If there exists an outgoing reaction $A^m_i \to B$ in $\mathcal{V}^m$ then we
    substitute it by the reaction $A^{m-1}_{i \tau_i} \to B$ with the same constant,
    i.e. outgoing reactions $A^m_i \to...$ are reattached to the heads of the
    limiting steps. Let us rearrange reactions from
    $\mathcal{V}^m$ of the form $B \to A^m_i$. These reactions have prototypes in
    $\mathcal{V}^{m-1}$ (before the last gluing). We simply restore these reactions.
    If there exists a reaction $A^m_i \to A^m_j$ then we find the prototype in
    $\mathcal{V}^{m-1}$, $A \to B$, and substitute the reaction by $A^{m-1}_{i
    \tau_i} \to B$ with the same constant, as for $A^m_i \to A^m_j$.
\item
After the previous step is performed, the vertices set is $\mathcal{A}^{m-1}$,
    but the reaction set differs from the reactions of the network
    $\mathcal{V}^{m-1}$: the limiting steps of cycles are excluded and the outgoing
    reactions of glued cycles are included (reattached to the heads of the limiting
    steps). To make the next step, we select vertices of $\mathcal{A}^{m-1}$ that are
    glued cycles from $\mathcal{V}^{m-2}$, substitute these vertices by vertices of
    cycles, delete the limiting steps, attach outgoing reactions to the heads of the
    limiting steps, and for incoming reactions restore their prototypes from
    $\mathcal{V}^{m-2}$, and so on.
\end{itemize}

After all, we restore all the glued cycles, and construct an acyclic reaction
network on the set $\mathcal{A}$. This acyclic network approximates relaxation
of the initial network. We call this system the {\em dominant system}.

Note that the reduction algorithm does not need precise values of the constants.
It is enough to have an initial ordering of the constants. Then, the auxiliary
network is obtained only from this ordering. However, after a first iteration,
and if the initial network contains cycles, some of the exit constant are
renormalized and the new rate constants become monomials of the old ones.
In order to prune again, we need to compare these monomials. Monomials
of well separated constants are generically well separated \cite{gorbanradulescu08}. However, a
freedom remains on ordering these new monomials, leading to several possible
reduced acyclic digraphs, given an initial digraph with ordering of the
constants (Figure \ref{fig1} of the main text).

\subsection*
{\bf Algorithm 2 : reduction of nonlinear networks with separation}

This algorithm consists of the following procedures.

{\bf I. Identification of QSS species and QE reactions.}

There are two methods of identification, trajectory based, and tropicalization based. Presently we are
using the trajectory based method.

\begin{description}
\item[\bf Detect slaved species.] After generating trajectories $c(t)$ for $t \in I$,
    for each species compute the distances $\delta_i = sup_{t \in I}
    |log(c_i(t))-log(c_i^*(t))| $. Use k-means clustering to separate species into
    two groups, slaved (small values of $\delta$) and slow (large values of $\delta$)
    species.
\item [\bf Prune.] For each $P_i$ (polynomial rate) corresponding to slaved species,
    compute the pruned version $\tilde P_i$ by eliminating all monomials that are
    dominated by other monomials of $P_i$.
\item [\bf Identify QE reactions and QSS species.] Identify, in the structure of
    $\tilde P_i$ the forward
 and reverse rates of QE reactions. This step can be performed by recipes presented in \cite{soliman2010unique}.
 The slaved species not involved in QE reactions are QSS.
\end{description}

\noindent
{\bf II. Exploiting QSS conditions, pruning intermediate species, pooling reactions}

\begin{description}
\item[\bf Define subsets and matrices]
Given the set of QSS (intermediate) species $I$, one defines the set ${\mathcal R}_I$ of
reactions acting on them. The terminal species $T$, are the the other species,
different from $I$, on which act the reactions from ${\mathcal R}_I$.
Define two stoichiometric matrices $S^f$ and $S^T$.
$S^f$ defines the fast subsystem and has a number of lines equal to the
number of QSS species, and a number of columns equal to the number of reactions
${\mathcal R}_I$. $S^T$ contains the stoichiometries of the terminal species
for the same reactions ${\mathcal R}_I$.
Species $I$ will be pruned, and reactions ${\mathcal R}_I$  will be pooled.
\item[\bf Compute elementary modes (EMs)]
Compute elementary modes of nonzero terminal stoichiometry as minimal
solutions of $S^f \gamma = 0$, $S^T \gamma \neq 0$, the minimality being defined with respect
to the number of nonzero coefficients.
$S^T \gamma \neq 0$ on the output of elementary modes packages such as METATOOL.
If the terminal stoichiometries of the EMs are dependent, restrict to a subset
of independent terminal stoichiometries.
\item[\bf Solve QSS equations]
Find approximate formal solutions for systems of QSS algebraic equations. This step is not
yet automatic. It will be automatized in subsequent work by using tropical geometry methods.
\item[\bf Find rates of EMs]
To each elementary mode $\gamma_i$, associate a kinetic law giving the rate of the EM as a
fonction of the terminal species concentrations $R_i^*(c_T)$.
Let $R(c_T)$ be the vector of rates of terminal species (the dependence on $c_T$
is direct, or indirect, via $c_I$ that can be now expressed as function of $c_T$) of
reactions in ${\mathcal R}_I$. Then the EM rates $R_i^*(c_T)$ must satisfy
$S^T R(c_T) = \sum R_i^*(c_T) S^T \gamma_i$. This equation has an
unique solution if the vectors $S^T \gamma_i$ are independent (this justifies the independence
condition for the terminal stoichiometries of EMs).
\end{description}

\noindent
{\bf III. Exploiting QE conditions, pruning QE reactions, pooling species}

\begin{description}
\item[\bf Define subsets and matrices]
Given the set of QE reactions $Q$, one defines the set $S$ of
species that are affected by them. The species $S$ are also affected
by other reactions that we call terminal, $Q_T$.
Define two stoichiometric matrices $S^f$ and $S^T$.
$S^f$ defines the fast subsystem and has a number of lines equal to the
cardinal of $E$, and a number of columns equal to the cardinal of $Q$.
$S^T$ contains the stoichiometries of the reactions reactions $Q_T$
for the same species $S$ (it has the same number of lines as $S^f$).
Reactions  $Q$ will be pruned and species $E$ will be pooled.
\item[\bf Compute species pools]
Species pools are computed as minimal solutions of $b S^f = 0$, $b S^T \neq 0$ (the
second condition stands for looking for conservation laws of the fast subsystem that
are not conserved by the entire network; the minimality condition means that we
compute elementary modes of the transpose matrix $S^f$).
\item[\bf Solve QE equations]
Same methods as for QSS conditions. Solve the QSS equations together with the conservation
of pools and express the concentrations of the species $E$ as functions of the pools  $c^*_i = b_i c$.
\item[\bf Find new rates]
Re-express (by substitution) the rate of each reaction from $Q_T$ in terms of
pools  $c^*_i = b_i c$.
\end{description}

\end{document}